\documentclass[twocolumn,prl,showpacs,byrevtex]{revtex4}
\usepackage{epsf,graphicx,amssymb}

\begin{document}
\title{Axisymmetric solitary waves on the surface of a ferrofluid}
\author{E. Bourdin}
\author{J.-C. Bacri}
\author{E. Falcon}
\email[Corresponding author]{}
\email[E-mail: ]{eric.falcon@univ-paris-diderot.fr}
\affiliation{Laboratoire Mati\`ere et Syst\`emes Complexes (MSC), Universit\'e Paris Diderot, CNRS (UMR 7057)\\10 rue A. Domon \& L. Duquet, 75 013 Paris, France}

\date{\today}

\begin{abstract}  
We report the first observation of axisymmetric solitary waves on the surface of a cylindrical magnetic fluid layer surrounding a current-carrying metallic tube. According to the ratio between the magnetic and capillary forces, both elevation and depression solitary waves are observed with profiles in good agreement with theoretical predictions based on the magnetic analogue of the Korteweg-deVries equation. We also report the first measurements of the velocity and the dispersion relation of axisymmetric linear waves propagating on the cylindrical ferrofluid layer that are found in good agreement with theoretical predictions.
\end{abstract}

\maketitle
Solitary waves or solitons are localized nonlinear waves that propagate almost without deformation 
due to the balance between the nonlinearity and the dispersion. Since the first observation of a solitary wave on the free-surface of water by Russel \cite{Russel1844}, and its interpretation using the Korteweg-de Vries equation (KdV) \cite{KdV1895}, it has been shown that the KdV equation describes a large class of solitons observed in various situations: acoustic waves on a crystal lattice, plasma waves, hydrodynamics internal or surface waves, elastic surface waves, and waves in optical fibers \cite{solitons}.  Most of them involve a localized elevation disturbance propagating within a quasi-one-dimensional plane system. Observations of axisymmetric solitary waves governed by the KdV equation are scarce~\cite{Lebovitch70}, and mainly concern waves in rotating fluids confined in a tube or on vortex lines. 
More recently, Bashtovoi {\em et al.} derived a KdV equation with an axisymmetric solitary waves solution propagating on the surface of a cylindrical magnetic fluid layer submitted to a magnetic field \cite{Bashtovoi83,Allemand}.  Without gravity, the stability of the cylindrical magnetic fluid layer is governed by the ratio between the magnetic force and the capillary one. According to its ratio, axisymmetric elevation (hump-like) or depression (hole-like) solitary waves are predicted with a subsonic or supersonic velocity \cite{Bashtovoi83,Allemand}. To our knowledge, direct observation of axisymmetric magnetic solitary waves have never been reported.

In this Letter, we report the first observation of axisymmetric solitary waves on the surface of a cylindrical ferrofluid layer submitted to an azimuthal magnetic field. Depending on the strength of the field, elevation or depression solitary waves are observed on the ferrofluid surface. A ferrofluid is a stable suspension of nanometric magnetic particles diluted in a carrier liquid (water or oil) that responds to an external applied magnetic field \cite{Rosen,Cebers}. Although the solitary waves are damped by viscous dissipation, we have shown that they keep the self-similar profile given by the solution of the KdV equation on a propagation length larger than their typical scale. Moreover, we also report the first measurement of the velocity and dispersion relation of axisymmetric magnetic linear waves in this system in good agreement with the theoretical predictions \cite{RD}.

\begin{figure}[t!]
\centerline{
 \epsfxsize=85mm
  \epsffile{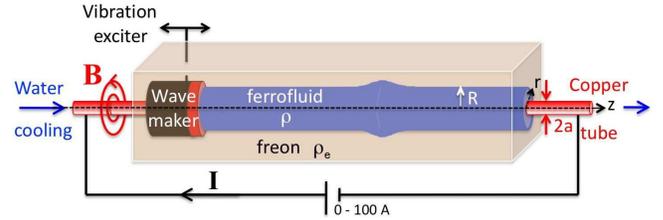}
}
\caption{(Color online) Experimental setup.}
\label{fig01}
\end{figure}
The experimental setup is shown in Fig.\ \ref{fig01}.  It consists of a cylindrical copper tube (50 cm in length, $a=1.5$ mm in outer radius and 0.5 mm in thickness) placed in the middle of a rectangular plexiglas container ($40 \times 40$ mm$^2$ side and 30 cm length) crossing both container end-sides in the center through hollow waterproof screws.  A dc electrical current, $I$, in the range 0--100 A is applied to the cylindrical conductor by means of a power supply. The current generates circular magnetic field lines around the tube with a radial decreasing amplitude. The corresponding radial magnetic force stabilizes a ferrofluid layer of outer radius $2.9 \leq R \leq 5$ mm, around the tube.  For the results reported below, $R=3.8$ mm for linear waves and $R=3.3$ mm for solitary waves. The ferrofluid used is an ionic aqueous suspension synthesized with 12.4\% by volume of maghemite  particles (Fe$_2$0$_3$ ; 7 $\pm 0.3$ nm in diameter) \cite{Talbot}. The properties of this magnetic fluid are: density, $\rho=1534\pm 1$ kg/m$^3$, initial magnetic susceptibility, $\chi_i=0.75$, magnetic saturation $M_{sat}=36\times 10^{3}$ A/m, and estimated dynamic viscosity 1.4 $\times 10^{-3}$ Ns/m$^2$. To avoid gravitational effects, the whole container is filled with Freon (C$_2$Cl$_3$F$_3$), a nonmiscible transparent fluid with a density, $\rho_e=1581\pm 1$ kg/m$^3$, close to the ferrofluid one. The surface tension between the ferrofluid and freon is $\gamma=5.5\times 10^{-3}$ N/m. A water cooling inside the tube drains off the Joule dissipation of the current-carrying electrodes (their contact resistance being 13.8 m$\Omega$, that is a dissipated power of roughly 140 W for $I=100$ A). The metallic tube is tightened to avoid parasitic vibration. To wit, one end of the tube is threaded to fix it with a nut, and a small chuck is used at the other end to tighten it. Surface waves are generated on the ferrofluid surface by the horizontal motion of a concentrical plexiglas tube, 9 mm (resp. 3 mm) in outer (resp. inner) diameter driven by an electromagnetic vibration exciter. The wavemaker is driven sinusoidally (in a frequency range from 0.5 to 10 Hz with a maximal amplitude of 2 mm) to study linear waves, or impulsively (typical duration of 0.05 s) to study solitary waves. Note that the wavemaker end is made of copper in order to increase the ferrofluid wetting. Axisymmetric waves propagating on the cylindrical ferrofluid layer are visualized with a high-resolution camera (Pixelink $2208 \times 3000$ pixels) located above the container, and are recorded with a 25 Hz (resp. 44 Hz) sampling for linear (resp. solitary) waves.

The magnetic induction generated by the carrying-current tube is up to 30 G at 100 A at a distance $r=8$ mm from the tube axis, that is the $z$-axis of the ($r$, $\theta$, $z$) cylindrical coordinate system. The magnetic induction being orthoradial $\vec{B}$=[$B_r=0$, $B_\theta\equiv B$,  $B_z=0$], i.e. throughout tangential to the free surface, the Rosensweig magnetic surface instability is absent \cite{Rosen,RD2}.  $B$ is measured with a transverse Hall probe via a gaussmeter (Bell 5100) as a function of the current and the distance $r$ from the tube in agreement with the usual law $B=\mu_0 I/(2\pi r)$ for $r\geq a$ where $\mu_0=4\pi\times 10^{-7}$ H/m is the magnetic permeability of the vacuum. The corresponding magnetic body force, $\vec{F}_{mag}=- \mu_0\chi I^2/(4\pi^2r^3)\vec{e}_r$,  is radial towards the $z$-axis. Since gravity is negligible ($\rho_e \simeq \rho$), this magnetic force stabilizes a uniform axisymmetric layer of a magnetic fluid of outer radius $R$ as soon as the capillary force by volume $F_{cap} \sim \gamma/R^2$ is small enough. Both magnetic and capillary effects are then compared by the dimensionless magnetic Bond number ${\rm Bo_m} \equiv F_{mag}/F_{cap} = \mu_0\chi I^2/(4\pi^2\gamma R)$.
 
Assuming no gravity and a thin tube radius ($a \ll R$), the dispersion relation of inviscid axisymmetric linear waves propagating on a magnetic fluid surface reads \cite{RD,RD2}
\begin{equation}
\omega^2=\frac{\gamma}{\rho R^3}kR\left[{\rm Bo_m}-1+\left(kR\right)^2\right]\frac{I_1(kR)}{I_0(kR)}
\label{RD}
\end{equation}
where $\omega\equiv 2\pi f$ is the angular frequency and $k\equiv2\pi/\lambda$ the wavenumber, $I_n$ and $K_n$ being, respectively, the modified Bessel functions of first and second kind of order $n$ (their ratio being a positive increasing function of $k$). When ${\rm Bo_m} \leq 1$, the capillary effects are greater than the magnetic ones, and an instability occurs [$\omega^2\leq 0$ in Eq.\ (\ref{RD})]: the cylindrical ferrofluid layer is unstable  to disturbances whose wavelengths $\lambda \geq 2\pi R/\sqrt{{\rm Bo_m}-1}$, and breaks up into a string of connected drops \cite{RD,decompositionRuss}. This is the magnetic analogue of the surface-tension-driven Rayleigh-Plateau instability when a thin cylindrical jet of a usual fluid breaks into a set of drops \cite{RP}.  When ${\rm Bo_m} > 1$, one has $\omega^2 > 0$ in Eq.\ (\ref{RD}): the cylindrical layer of ferrofluid is stable whatever the wavelength disturbance, and axisymmetric linear waves can propagate on its surface.

\begin{figure}[t!]
\centerline{
  \epsfxsize=75mm
  \epsffile{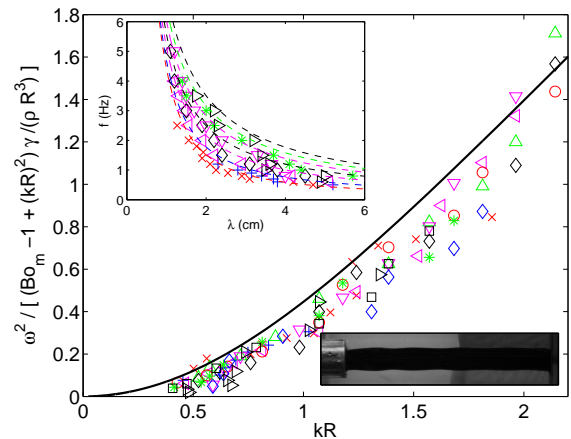}
}
\caption{(Color online) Dimensionless dispersion relation of linear cylindrical waves for various applied currents $I$ from 40 to 100 A corresponding to ${\rm Bo_m}=1.85$ ($\times$), 2.69 ($+$), 4.17 ($\triangleleft$), 4.89($\vartriangle$), 5.67 ($\lozenge$),  6.51 ($\circ$),  7.4 ($\triangledown$), 8.36 ($\lozenge$), 9.37 ($\ast$), 10.44 ($\square$), and 11.57 ($\triangleright$). Solid line corresponds to the theoretical prediction of Eq.\ (\ref{RD}). Top inset: Frequency $f$ as a function of the wavelength $\lambda$ for different $I$ with a 10 A step. Dashed lines are from Eq.\ (\ref{RD}). Bottom inset: Snapshot of linear waves ($f=3.5$ Hz, ${\rm Bo_m}=6.5$ - wavemaker is visible on the left-hand side - 10 cm size window). }
\label{fig02}
\end{figure}

We first measure the dispersion relation of such linear waves.  The wavemaker is driven sinusoidally in order to generate surface waves at the interface between the freon and the ferrofluid. A typical snapshot of such axisymmetric linear waves is shown in the bottom inset of Fig.\ \ref{fig02}. The top inset of Fig.\ \ref{fig02} shows the wavelength, $\lambda$, of surface waves in response to the forcing frequency, $f$, for different applied currents, $I$, that is for different $ {\rm Bo_m} \sim I^2 $ ranging from 1 to 12. $\lambda$ is found to decrease with increasing frequency whatever ${\rm Bo_m}$. When expressed in the rescaled variables $\omega^2/\left[({\rm Bo_m}-1+\left(kR)^2\right)\gamma/(\rho R^3)\right]$ and $kR$, all these data collapse on one single master curve (solid line) predicted by Eq.\ (\ref{RD}) (see Fig.\ \ref{fig02}). Note that no adjustable parameter is used when comparing the data and the theoretical dispersion relation of axisymmetric magnetic surface waves.

\begin{figure}[t!]
\centerline{
\epsfxsize=80mm
\epsffile{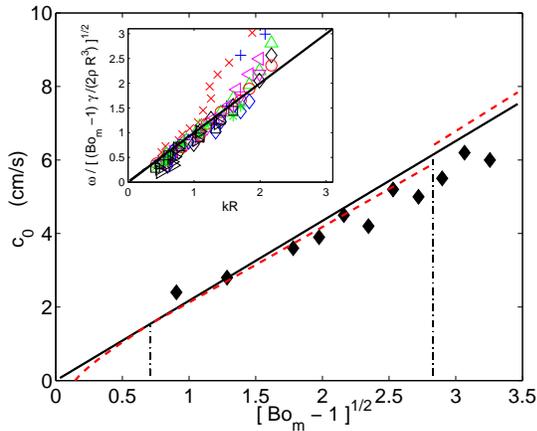} 
}
\caption{(Color online) Velocity of linear axisymmetric magnetic waves as a function of rescaled Bond number. Experimental ($\blacklozenge$) and theoretical [solid line of Eq.\ (\ref{RDapprox})]  linear wave velocity $c_0$. Dashed line: theoretical solitary wave velocity $c$ of Eq.\ (\ref{cs}) with $A_0=0.5$ mm, $R=3.8$ mm (see text). Dash-dotted lines correspond to ${\rm Bo_m}= 3/2$ and 9. Inset: Rescaled dispersion relation $\omega/(c_0/R)$ vs. $kR$. Same symbols as Fig.\ ~\ref{fig02}. Solid line has a slope 1.}
\label{fig03}
\end{figure}

Using the expansion of the modified Bessel functions $\frac{I_1(x)}{I_0(x)}\sim\frac{x}{2}-\frac{x^3}{16}$ \cite{HandbookMath}, the dispersion relation of Eq.\ (\ref{RD}) in a long-wavelength limit ($kR\ll 1$) reads \cite{Allemand}
\begin{equation}
\omega=c_0k\left[1-\frac{1}{16}\frac{{\rm Bo_m}-9}{{\rm Bo_m}-1}k^2R^2\right]{\rm , \ } c_0=\sqrt{\frac{\gamma}{\rho R}}\sqrt{\frac{{\rm Bo_m}-1}{2}}
\label{RDapprox}
\end{equation}
$c_0$ being the velocity of linear waves for ${\rm Bo_m} > 1$. Note that both $c_0$ and the sign of the dispersive term $\sim k^3$ in Eq.Ê(\ref{RDapprox}) depend on ${\rm Bo_m}$. In order to extract the velocity of linear waves, the inset of Fig.\ \ref{fig03} displays the previous data in variables $\omega/(c_0/R)$ and $kR$. As expected, for small $kR$, all the data collapse on a single linear curve of slope 1. Note that for larger $kR$ and for ${\rm Bo_m}\simeq 1$, a departure from the prediction of the linear term of Eq.\ (\ref{RDapprox}) is observed since the dispersive effects become important [{\em i.e.}, the $k^2/({\rm Bo_m}-1)$ term in Eq.\ (\ref{RDapprox})]. For $kR\ll 1$, the slope of each curve $\omega$ vs. $k$ thus gives, for each ${\rm Bo_m}$ value, a direct measurement of the velocity of linear waves. These values are plotted in Fig.\ \ref{fig03} and are found in rough agreement with the theoretical velocity of Eq.\ (\ref{RDapprox}) with no adjustable parameter. To our knowledge this is the first measurement of the velocity and dispersion relation of axisymmetric magnetic linear waves on the surface of a ferrofluid.

Let us now focus on axisymmetric magnetic solitary waves. First, let us assume no viscosity and no gravity. In the long-wavelength limit ($kR\ll 1$), the dispersion is small and the linear wave velocity is $c_0$. When the interface deflection $A(z,t)$ is also small, such that nonlinear effects have the same order of magnitude as dispersive ones, it is governed at the leading order by a magnetic analogue of the Korteweg-de Vries equation \cite{Bashtovoi83,Allemand}

\begin{equation}
A_t+c_0A_z+\alpha AA_z +\beta A_{zzz}=0 {\rm ,}
\label{KdV}
\end{equation}

with $\alpha=\frac{2{\rm Bo_m}-3}{2^{3/2}\sqrt{{\rm Bo_m}-1}}\sqrt{\frac{\gamma}{\rho R^3}}$ the nonlinear coefficient, and $\beta=\frac{{\rm Bo_m}-9}{2^{9/2}\sqrt{{\rm Bo_m}-1}}\sqrt{\frac{R^3 \gamma}{\rho}}$ the dispersive coefficient, and ${\rm Bo_m}>1$. The axisymmetric magnetic solitary wave solution of Eq.\ (\ref{KdV}) reads \cite{Bashtovoi83,Allemand}

\begin{equation}
A(z,t)=A_0 {\rm sech}^2\left(\frac{z-ct}{L}\right) {\rm , \ } L=\sqrt{\frac{3R^3}{2A_0}\frac{{\rm Bo_m}-9}{2{\rm Bo_m}-3}} {\rm ,}
\label{solKdV}
\end{equation}
with $c$ the velocity of solitary wave
\begin{equation}
c=c_0\left(1+\frac{A_0}{6R}\frac{2{\rm Bo_m}-3}{{\rm Bo_m}-1}\right) {\rm ,}
\label{cs}
\end{equation} 
and $L$ is the length scale of the solitary wave. Equations (\ref{solKdV}) and (\ref{cs}) show that there exists a continuous family of soliton solutions with parameter $A_0$ (the extremum amplitude of the wave). Since $\alpha$, $\beta$, $c$ and $L$ depend on ${\rm Bo_m}$, an elevation ($A_0>0$) or depression ($A_0<0$) solitary wave is predicted that propagates with a supersonic ($c>c_0$) or subsonic ($c<c_0$) speed. All the possible solutions are summarized in Tab.\ \ref{tableau}. 

\begin{table}[h!]
\begin{center}
\begin{tabular}{cccccc}

\hline
\hline
${\rm Bo_m}$ & $\alpha$ & $\beta$ & $c$ & $A_0$ & solitary wave \\
\hline
$1<{\rm Bo_m}<3/2$ & - & - & $<c_0$ & + & subsonic elevation\\
$3/2<{\rm Bo_m}<9$ & + & - & $<c_0$ & - & subsonic depression\\
${\rm Bo_m}>9$ & + & + & $>c_0$ & + & supersonic elevation\\
\hline
\hline

\end{tabular}
\end{center}
\caption{Properties of axisymmetric solitary wave solution of Eqs.\ (\ref{solKdV}) and (\ref{cs}) according to the magnetic Bond number.}
\label{tableau}
\end{table}

\begin{figure}[Ht!]
\centerline{
\begin{tabular}{c}
  \epsfxsize=75mm
  \epsffile{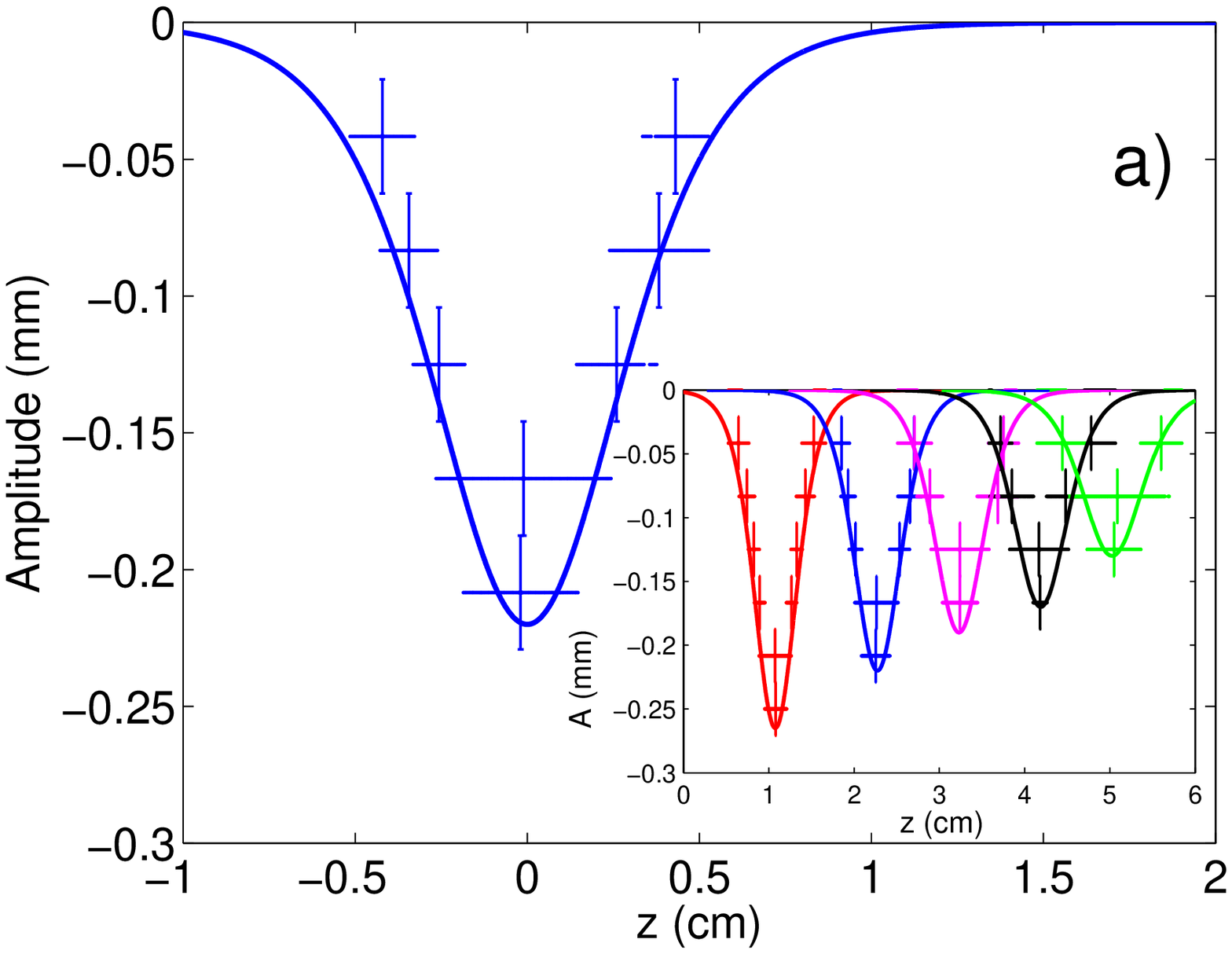}\\
  \epsfxsize=75mm
    \epsffile{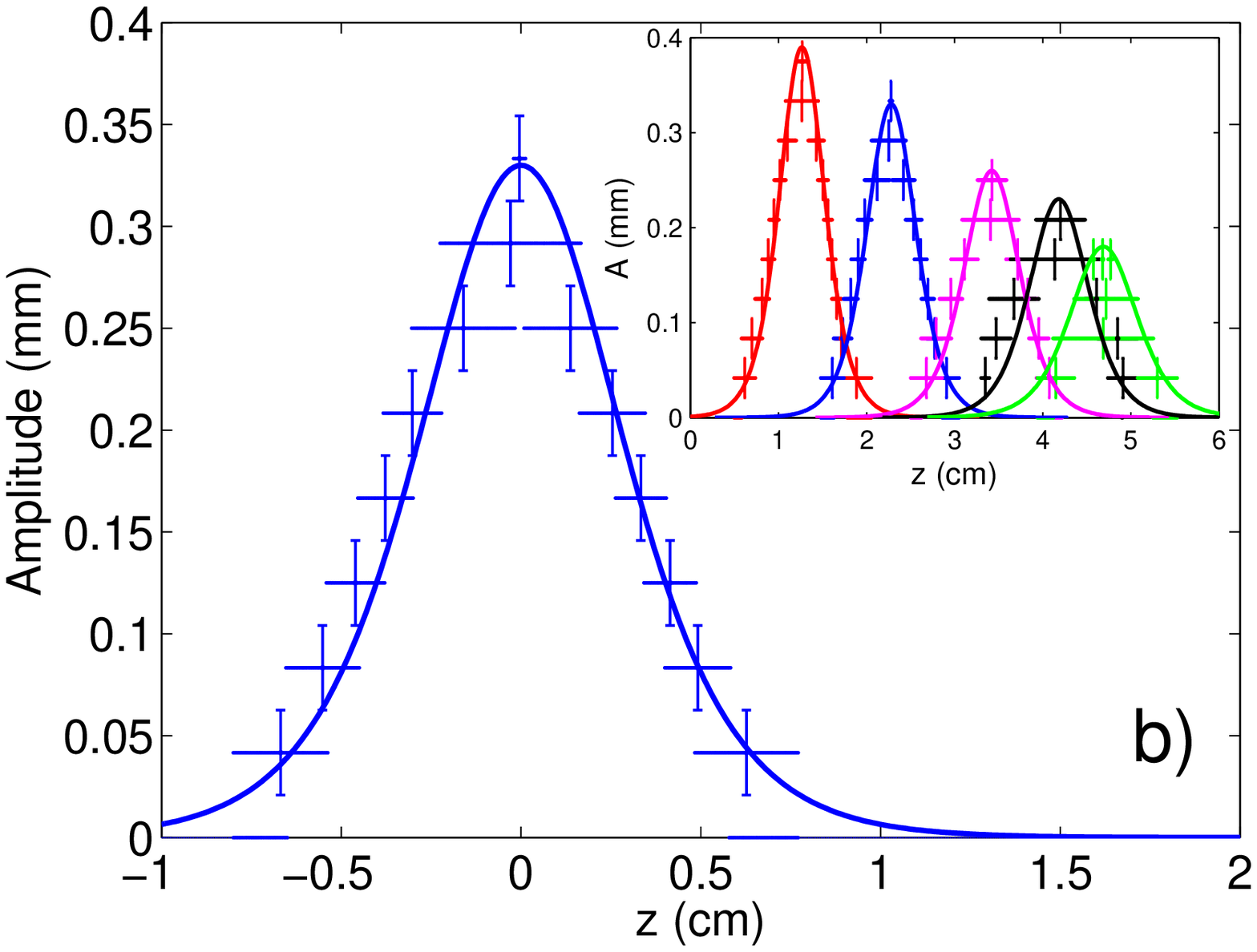}
\end{tabular}
}
  \caption{(Color online) {\bf (a)} Axisymmetric profile of a depression magnetic solitary wave for ${\rm Bo_m}=8.3$ ($I= 80$ A), $t=0.33$ s, {\bf (b)} elevation magnetic solitary wave for ${\rm Bo_m}=10.5$ ($I=90$ A), $t=0.32$ s, centered on its extremum. Solid lines are the theoretical profiles of KdV solitons derived from Eq.\ (\ref{solKdV}) with no adjustable parameter. Insets: Profiles of solitary waves at different times [(a): 0.16, 0.33, 0.47, 0.62, and 0.74 s; (b): 0.17, 0.32, 0.47, 0.58, and 0.65 s] during its propagation over 10 times its typical size $L\sim 4$ mm. Initial amplitudes $A_0=\pm 0.5$ mm. $R=3.3$ mm. $z$-axis origin is located on the wavemaker. }
  \label{fig05}
  \end{figure}
  
We have performed a study of axisymmetric solitary waves on the surface of the ferrofluid layer ($R=3.3$ mm) around a copper tube carrying current in the range 60 to 110 A ($4 \leq {\rm Bo_m} \leq 14$). We impulsively drive the shaker to generate solitary waves: the wavemaker is pushed forward to generate a pulse on the fluid interface leading to either an elevation or a depression pulse according to the value of ${\rm Bo_m}$. The interface deflection $A(z,t)$ is detected from the images recorded by the camera using standard \textsc{ImageJ} binarization and edge detection processes. The profile is displayed  in Fig.\ \ref{fig05}a for a depression pulse (${\rm Bo_m}=8.3$) and in Fig.\ \ref{fig05}b for an elevation pulse (${\rm Bo_m}=10.5$). Both recordings are in good agreement with the profiles of elevation and depression KdV solitary waves given by Eq. (\ref{solKdV}). Note that once $A_0$ is known the theoretical profile as well as the velocity of the solitary wave given by Eqs. (\ref{solKdV}) and (\ref{cs}) do not involve any adjustable parameter. Those isolated pulses involve typical amplitude and size that are in the range of validity required for the derivation of Eq. (\ref{KdV}), that is,  corresponding to small dispersion ($L^2\gg R^2$), and small nonlinearities ($|A_0|\ll R$), both of same order of magnitude ($R^3\sim |A_0| L^2$). Note that no solitary wave has been observed for ${\rm Bo_m}<4$ since its predicted amplitude ($A_0 \sim$ few mm for $L\sim 1$ cm) is too large compared to our tube radius $R$ to have small nonlinearities. Inset of Fig.\ \ref{fig05}a (resp. Fig.\ \ref{fig05}b) shows the profile of the depression  (resp. elevation) pulse recorded at different times corresponding to a total propagation distance up to 10 times its typical size. The recorded profiles are in good agreement with the KdV magnetic solitary wave all along the propagation. Note, however, that for farther distances the cumulative effect of dissipation leads to small amplitudes that are hardly measurable by the camera (0.04 mm/pixel). For both the elevation and depression solitary waves, dissipation leads to an extremum amplitude $A_0(z)$ that decreases linearly with the propagation distance $z$. By rescaling all the profiles displayed in each inset of Fig.\ \ref{fig05} with the variables $|A(z)/A_0(z)|$ vs. $z/L$, all the data lie on a single curve predicted by Eq. (\ref{solKdV}) (not shown here). This means that the pulse keeps a self-similar shape over a distance up to 10 times its typical size and is in a good agreement with the profile derived from KdV magnetic equation.  Finally, the solitary wave velocity, $c$, is measured all along its propagation by the successive locations of amplitude extrema, $|A_0(z)|$, at different times. We find that $c \simeq c_0$ ($\sim$ few cm/s) with a dependence on ${\rm Bo_m}$ roughly comparable to that predicted by Eq.\ (\ref{RDapprox}). Note that $c$ is predicted by Eq.\ (\ref{cs}) to slightly depend on $A_0$ with a correction with respect to $c_0$ up to 10\% when ${\rm Bo_m}\sim 10$ (see dotted line on Fig.\ \ref{fig03}). Since our velocity measurement accuracy is 6\%, we cannot thus discriminate from a subsonic to a supersonic solitary wave as predicted in Table I (see also Fig.\ \ref{fig03}).

In conclusion, we have reported the first observation of depression and elevation axisymmetric solitary waves on the surface of a cylindrical magnetic fluid layer and found that their shapes are in good agreement with the ones predicted from the axisymmetric KdV solitary wave solutions. A possible extension of this work would be the study of the collisions between these new solitary waves.


\begin{acknowledgments}
We thank D. Talbot, S. H\'enon, J. S\'ebilleau, A. Lantheaume and C. Laroche. This work has been supported by ANR Turbonde BLAN07-3-197846.
\end{acknowledgments}

\end{document}